\documentclass{article}

\usepackage{arxiv}
\usepackage{algorithm2e}
\usepackage[utf8]{inputenc} 
\usepackage[T1]{fontenc}    
\usepackage{hyperref}       
\usepackage{url}            
\usepackage{booktabs}       
\usepackage{amsfonts}       
\usepackage{nicefrac}       
\usepackage{microtype}      
\usepackage{lipsum}		
\usepackage{graphicx}
\usepackage{natbib}
\usepackage{doi}

\title{A Simulated Annealing Approach to Identical Parallel Machine Scheduling}

\author{ \hspace{1mm}Jiaxing Li \\
	Piedmont Hills High School\\
	San Jose, CA 95132 \\
	\texttt{lijiaxing214@proton.me} \\
	\And
	\hspace{1mm}David Perkins \\
	Department of Computer Science\\
	Colgate University\\
	Hamilton, NY 13346 \\
	\texttt{dperkins@colgate.edu} \\
}

\hypersetup{
pdftitle={A Simulated Annealing Approach to Identical Parallel Machine Scheduling},
pdfsubject={},
pdfauthor={Jiaxing Li, David Perkins},
pdfkeywords={},
}

\begin{document}
\maketitle

\begin{abstract}
This paper studies the application of the simulated annealing metaheuristic on the identical parallel machine scheduling problem, a variant of the broader optimal job scheduling problem. In the identical parallel machine scheduling problem, $n$ jobs are to be assigned among $m$ machines. Furthermore, each job takes a certain amount of time that remains constant across machines. The goal of the paper is to schedule $n$ jobs on $m$ machines and minimize the maximum runtime of all machines. Both exact and heuristic methods have been applied to the problem, and the proposed algorithm falls in the heuristic category, making use of the simulated annealing metaheuristic. Compared to exact algorithms, simulated annealing was found to yield near-optimal solutions in comparable or less time for all problem cases.
\end{abstract}


\section{Introduction}
Despite being simple conceptually, identical machine scheduling has extensive applications in real-world scenarios where jobs take constant time, can be done on any "machine," and do not have precedence relationships. For example, consider a 3D printing shop consisting of $m$ printers tasked to build a set of $n$ orders. On identical printers, each order takes the same time to complete. In addition, every order can be built on any printer and the sequence the orders are built does not matter. In this example, the orders are jobs, and printers are machines. The runtime of each machine is the time taken to finish all jobs assigned to it. The goal is to minimize the maximum runtime of all printers, since the machine with the maximum runtime will be the bottleneck. In optimal job scheduling, the maximum runtime of all machines is called $C_{max}$, or the \textit{makespan}. The simulated annealing metaheuristic is a simple yet fast method to solve the identical machine scheduling problem, producing near-optimal solutions rapidly. Morever, this metaheuristic is simple to understand and implement. This metaheuristic, along with the heuristic, is discussed in section \ref{sec:proposed}. The results of computational experiments comparing the proposed algorithm with others are then given in section \ref{sec:compexperiments} and discussed in \ref{sec:discussion}.

\section{Literature Review}
\label{sec:litreview}
The parallel machine scheduling problem has been extensively studied. Minimizing the $mean$ makespan can be done in O($n$ log $n$) time complexity, with the SPT (shortest processing time first) algorithm. However, the goal of this paper, minimizing $C_{max}$, was proven to be NP-Complete by \cite{Bernstein1989-qx} via the 3-partition problem. Thus, while exact algorithms exist, they are only applicable in small to medium cases.

\subsection{Exact Methods}
\label{sec:exact}
The most common exact methods used to solve the identical parallel machine scheduling problem include branch-and-bound, dynamic programming, and integer programming. \cite{Mokotoff2004-ti} investigated the use of the cutting-plane algorithm, an integer programming method. \cite{DellAmico1995-rv} used the branch and bound method, finding that it performed better than dynamic programming on average. Furthermore, a follow-up paper by \cite{DellAmico2005-bw} showed this algorithm was better than the integer programming approach of \cite{Mokotoff2004-ti}. \cite{Chen2020-jm} explored both exact and heuristic methods to solve the problem. The exact methods considered by \cite{Chen2020-jm} were based on branch and bound. Notably, dynamic programming was incorporated as an upper bounding technique to assist the algorithm. Computation experiments revealed the impressive efficiency increase this strategy provided. On average, the branch and bound method with dynamic programming explored about only one percent of the solution space explored by classic branch and bound. 

\subsection{Heuristic Methods}
\label{sec:heuristic}
Heuristic, or approximate, methods range from simple ones such as greedy algorithms to complex metaheuristics such as genetic algorithms. The nature of heuristic methods requires the consideration of both run time and approximation factor. Approximation factor quantifies how close the solution the heuristic yields is to optimal, and is $\frac{t_{h}}{t_{o}}$ where $t_{h}$ is the value of $C_{max}$ reached by the heuristic method, and $t_{o}$ is the optimal value of $C_{max}$. Approximation factor is either an inherent property of the algorithm and calculated via mathematical analysis or a target value set by the user, where the algorithm runs until it reaches that value. The proposed algorithm uses the latter approach. One of the simplest algorithms is list scheduling, which greedily assigns jobs to the earliest available machine. List scheduling was proven by \cite{Graham1969-un} to have an approximation factor of $2 - \frac{1}{m}$. The algorithm developed by \cite{Huang2021-ig} has an approximation factor of $\frac{11}{9}$ with time complexity O($m$ log $m$ + $n$). This algorithm uses an approach similar to the FFD (First Fit Decreasing) algorithm for bin packing. Moving up in complexity, \cite{ANGHINOLFI2021416} used an established heuristic, constructive heuristic, along with a novel local search method called exchange search. Notably, their algorithm optimizes both $C_{max}$ and the varying energy cost to run the machines. \cite{Lee2021-ww} explore the problem while considering setup times, the time needed to set up a machine, and the ability to split jobs among different machines. Compared to genetic algorithms, their setup time-based iterative algorithm (STIA) algorithm performed better, although the authors do note that the performance of genetic algorithms is inherently worse when jobs can be split. 

\section{Proposed Algorithm}
\label{sec:proposed}
The proposed algorithm combines a conceptually simple heuristic with a metaheuristic, simulated annealing, to ensure efficient performance. The algorithm is divided into two main phases: the initialization phase and the metaheuristic phase. In the initialization phase, jobs are randomly assigned to machines to create an initial solution. In the metaheuristic phase, the simulated annealing metaheuristic is used to guide a heuristic to improve the solution towards the optimal solution. Therefore, this algorithm does not immediately construct an optimal solution, but rather generates a solution that is likely suboptimal and steadily improves it until it becomes optimal.

\subsection{Heuristic}
\label{sec:proposedheuristic}
The heuristic procedure employs two fundamental strategies, moving and swapping jobs, to minimize $C_{max}$. For both strategies, the heuristic selects two machines. When moving, a job is selected from one machine and moved to the other. When swapping, one job is selected from each machine and interchanged. Furthermore, this heuristic algorithm selects both machines and jobs randomly. The heuristic is as follows:
\begin{algorithm}[H]
    \SetAlgoLined
    \SetKwInOut{KwIn}{Input}
    \SetKwInOut{KwOut}{Output}
    \KwIn{A set of $m$ machines with $n$ jobs distributed among them, and a floating-point number $0 < r < 1$ }
    \KwOut{A set of $m$ machines with $n$ jobs distributed among them}
    $a \leftarrow $ random(0, 1) \\
    $m_{1} \leftarrow$  a random machine \\
    $m_{2} \leftarrow$  a random machine that is not $m_{1}$ \\
    \eIf{$a > r$} {
        $j_{1} \leftarrow$  random job from $m_{1}$ \\
        $j_{2} \leftarrow$  random job from $m_{2}$ \\
        Swap $j_{1}$ and $j_{2}$ between $m_{1}$ and $m_{2}$ \\
    } {
        $j_{1} \leftarrow$  random job from $m_{1}$ \\
        Move $j_{1}$ from $m_{1}$ to $m_{2}$ \\
    }
    \caption{Heuristic Procedure for Job Scheduling}
    \label{alg:heuristic}
\end{algorithm}    

\subsection{Metaheuristic}
\label{sec:proposedmetaheuristic}
The function of a metaheuristic is to guide a heuristic towards a good solution. The metaheuristic used in this paper is simulated annealing. Simulated annealing was first introduced by \cite{Kirkpatrick1983-gu} in 1983, who used it to solve the traveling salesman problem. Simulated annealing is inspired by the annealing process in metallurgy, a process relying on heating and cooling metal to modify properties such as ductility or hardness. In this metaheuristic, an initial "temperature" is set and decays exponentially. If the heuristic yields a solution worse than the previous, the probability of accepting that solution is proportional to the temperature. Thus, over many iterations, the metaheuristic will gradually guide the heuristic towards progressively better moves. In addition, this technique allows the heuristic to explore bad initial moves that may lead to good solutions. The simulated annealing procedure used in this paper is as follows:
\begin{algorithm}[H]
    \SetAlgoLined
    \SetKwInOut{KwIn}{Input}
    \SetKwInOut{KwOut}{Output}
    \KwIn{A set of $m$ machines with $n$ jobs randomly distributed among them, an estimated optimal value $k$ of $C_{max}$, a desired approximation factor $p > 1$, an initial temperature $T$, a decay rate $0 < D < 1$, a parameter $a > 0$, and a maximum elapsed time $M$}
    \KwOut{A set of $m$ machines with $n$ jobs distributed among them such that $C_{max} \leq k \cdot p$}
    \While{$t < M$} {
        \If{$C_{max} \leq k \cdot p$} {
            Return $m$ \\
        }
        Call the heuristic with the current solution $m$ to produce a neighboring solution $x$ \\
        \eIf{$C_{max}(x) > C_{max}(m)$} {
            $r \leftarrow$ random(0, 1) \\
            \eIf{$r < exp(\frac{-a}{T})$} {
                Accept $x$ as the new solution and assign it to $m$\\
            } {
                Reject $x$ and keep $m$ as the new solution \\
            }
        } {
            Accept $x$ as the new solution and assign it to $m$ \\
        }
        $T \leftarrow T \cdot D$
    }
    Return $m$ \\
    \caption{Simulated Annealing Procedure}
    \label{alg:simannealing}
\end{algorithm}   
To estimate the optimal value $k$ of $C_{max}$, the following formula is used where $l$ is the length of the longest job, $s$ is the sum of the length of all jobs, and $n$ is the number of jobs: 

\(k = max(l, \frac{s}{n})\)

However, as an estimate, it cannot be guaranteed that this optimal value can be reached, nor that it can even be reached within the approximation factor $p$. In situations where the optimal $C_{max}$ is greater than $k \cdot p$, the performance of the algorithm decays to requiring $M$ time, which was observed to happen most often if the ratio of $n$ to $m$ was small. Thus, a value of $M$ must be derived that can both end the algorithm early if $k \cdot p$ is unreachable while also giving it enough time to explore the solution space. From the observed test cases, it was found that a good value of $M$ can be given by the number of jobs $n$ taken in milliseconds. Nevertheless, this observation should not be taken as a rule that holds true in all cases.

\section{Computational Experiments}
\label{sec:compexperiments}
The proposed algorithm was coded in Java. Testing was done on a computer running Ubuntu 22.04 LTS with an Intel Core i7-6820HQ processor using the 64-bit Oracle JDK 22. The problem sets used were obtained by generating jobs with random lengths between 1 and 100. The number of machines and jobs in each problem is denoted by $m$ and $n$ respectively. The approximation factor $p$ used was 1.01, equivalent to finding solutions within 1\% of optimal. The parameters $T$, $D$, and $a$ for simulated annealing were 10, 0.9, and 1, respectively. The average time required by the algorithm over ten instances of each problem is given in seconds. Results of the proposed algorithm are denoted by PA while the results of the cutting-plane method of \cite{Mokotoff2004-ti} are denoted by CP. 
\begin{table}
    \centering
    \begin{tabular}{cccc}
           \hline
  $m$ & $n$ & PA & CP \\ 
  \hline
  3  & 5 & 0.004 & 0.0041\\ 
  \hline
  3 & 6 & 0.004 & 0.005\\ 
  \hline
  3 & 7 & 0.005 & 0.0053\\ 
  \hline
  3 & 8 & 0.007 & 0.0059\\ 
  \hline
  3 & 9 & 0.006 & 0.0074\\ 
  \hline
  3 & 10 & 0.006 & 0.0055\\ 
  \hline
  3 & 11 & 0.005 & 0.0048\\ 
  \hline
  3 & 12 & 0.004 & 0.0021\\ 
  \hline
  3 & 13 & 0.004 & 0.0021\\ 
  \hline
  3 & 14 & 0.002 & 0.0011\\ 
  \hline
  3 & 15 & 0.002 & 0.0025\\ 
  \hline
  4 & 5 & 0.001 & 0.0029\\ 
  \hline
  4 & 6 & 0.003 & 0.0032\\ 
  \hline
  4 & 7 & 0.003 & 0.0057\\ 
  \hline
  4 & 8 & 0.005 & 0.0072\\ 
  \hline
  4 & 9 & 0.007 & 0.0072\\ 
  \hline
  4 & 10 & 0.009 & 0.0084\\ 
  \hline
  4 & 11 & 0.009 & 0.0078\\ 
  \hline
  4 & 12 & 0.010 & 0.0082\\ 
  \hline
  4 & 13 & 0.009 & 0.0088\\ 
  \hline
  4 & 14 & 0.003 & 0.0084\\ 
  \hline
  4 & 15 & 0.004 & 0.007\\ 
  \hline
  5 & 6 & 0.001 & 0.0028\\ 
  \hline
  5 & 7 & 0.002 & 0.0031\\ 
  \hline
  5 & 8 & 0.004 & 0.0056\\ 
  \hline
  5 & 9 & 0.007 & 0.0096\\ 
  \hline
  5 & 10 & 0.009 & 0.011\\ 
  \hline
  5 & 11 & 0.010 & 0.012\\ 
  \hline
  5 & 12 & 0.009 & 0.011\\ 
  \hline
  5 & 13 & 0.011 & 0.012\\ 
  \hline
  5 & 14 & 0.012 & 0.013\\ 
  \hline
  5 & 15 & 0.011 & 0.02\\ 
  \hline
  100 & 200 & 0.187 & 6.6875\\ 
  \hline
  100 & 500 & 0.087 & 6.9016\\ 
  \hline
  100 & 1000 & 0.075 & 14.725\\ 
  \hline

    \end{tabular}
    \caption{Computational Experiment Results}
    \label{tab:results}
\end{table}
\section{Discussion}
\label{sec:discussion}
As shown by the results, simulated annealing performed well in all test cases compared to the cutting-plane method, with the most dramatic improvements exhibited in the largest test cases. While heuristic algorithms cannot guarantee the optimality of a solution, simulated annealing was able to get within at least 1\% of optimal in approximately the same time for smaller test cases and significantly less time for larger ones. Therefore, the proposed algorithm is suitable for larger test cases where time is a priority, while it does not offer a significant advantage in smaller test cases.  

\section{Conclusion}
In this paper, the simulated annealing metaheuristic was employed to solve the identical parallel machine scheduling problem. The results show that this method is viable for both large and small problem sets. Despite this success, there are many areas for improvement. First, since simulated annealing can only improve an initial solution, the better the initial solution is, the better the algorithm will perform. Thus instead of randomly distributing the jobs, a simple heuristic method such as the list scheduling method mentioned in \ref{sec:heuristic} could be implemented in the initialization stage before the metaheuristic is even invoked. Moreover, as discussed in \ref{sec:proposedmetaheuristic}, the current method for estimating the optimal value of $C_{max}$ has its flaws and could be improved in order to avoid unattainable values. Also, improvements such as tuning the simulated annealing parameters can be made to further increase efficiency. In the future, simulated annealing could be applied to related problems, such as the flow-shop problem where jobs must be scheduled in a given order. In addition, to more closely replicate real-world situations the identical parallel machine scheduling problem could be modified to optimize factors in conjunction with makespan such as energy cost, lateness, and many others. While these additions do dramatically increase the complexity of the problem, the use of metaheuristics such as simulated annealing lead to a more dramatic speedup than it provides on simpler problems such as the one discussed in this paper. 

\bibliographystyle{unsrtnat}
\bibliography{references}  






\end{document}